\documentclass{aastex}
\usepackage{emulateapj5}

\newcommand{\msun}{\>{\rm M_{\odot}}}

\newcommand{\bdm}{\begin{displaymath}}
\newcommand{\edm}{\end{displaymath}}
\newcommand{\beq}{\begin{equation}}
\newcommand{\eeq}{\end{equation}}
\newcommand{\bit}{\begin{itemize}}
\newcommand{\eit}{\end{itemize}}
\newcommand{\ben}{\begin{enumerate}}
\newcommand{\een}{\end{enumerate}}
\newcommand{\bfi}{\begin{figure}[htb]}
\newcommand{\bpfi}{\begin{figure}[p]}

\shorttitle{Fueling and Feedback in IC342's Nucleus}
\shortauthors{Schinnerer et al.}

\begin{document}

\title{Self-Regulated Fueling of Galaxy Centers:\\
       Evidence for Star-Formation Feedback in IC342's Nucleus\altaffilmark{1}}

\author{Eva Schinnerer\altaffilmark{2},
Torsten B\"oker\altaffilmark{3},
David S. Meier\altaffilmark{4,5},
Daniela Calzetti\altaffilmark{6}}
\altaffiltext{1}{Based on observations carried out with the IRAM
  Plateau de Bure Interferometer. IRAM is supported by INSU/CNRS
  (France), MPG (Germany) and IGN (Spain).}

\altaffiltext{2}{Max-Planck-Institut f\"ur Astronomie, K\"onigstuhl 17,
    D-69117 Heidelberg, Germany}
\altaffiltext{3}{European Space Agency, Dept. RSSD, Keplerlaan 1, 
    2200 AG Noordwijk, Netherlands}
\altaffiltext{4}{David S. Meier is a Jansky Fellow of the National 
    Radio Astonomy Observatory}
\altaffiltext{5}{National Radio Astronomy Observatory, P.O. Box O, 
    Socorro, NM 87801, U.S.A.}
\altaffiltext{6}{Department of Astronomy, University of Massachusetts, 
    Amherst, MA 01003, U.S.A.}


\begin{abstract}
  Using new, high-resolution interferometric observations of the CO
  and HCN molecules, we directly compare the molecular and ionized
  components of the interstellar medium in the center of the nearby
  spiral galaxy IC\,342, on spatial scales of $\approx$ 10\,pc. The
  morphology of the tracers suggests that the molecular gas flow
  caused by a large-scale stellar bar has been strongly affected by
  the mechanical feedback from recent star formation activity within
  the central 100\,pc in the nucleus of the galaxy. Possibly, stellar
  winds and/or supernova shocks originating in the nuclear star
  cluster have compressed, and likely pushed outward, the infalling
  molecular gas, thus significantly reducing the gas supply to the
  central 10\,pc. Although our analysis currently lacks kinematic
  confirmation due to the face-on orientation of IC\,342, the
  described scenario is supported by the generally observed repetitive
  nature of star formation in the nuclear star clusters of late-type
  spiral galaxies.
\end{abstract}


\section{Introduction}

Any ``activity'' found in a galactic nucleus -- be it due to massive
star formation, an accreting super-massive black hole, or a
combination of both -- requires the inflow of gas into the central few
pc. On scales of kpc, the gas flow is regulated by the response to
asymmetries in the gravitational potential, e.g. stellar bar or tidal
interactions \citep{sak99,she05}. Numerical models
\citep[e.g.][]{ath92,sel93} have demonstrated that large-scale stellar
bars are especially efficient in moving gas into the central
kpc. However, no clear picture is yet available for the gas flow
inside a few 100 pc from the nucleus \citep{gar03,wad04}.

Arguably the simplest galaxies to observationally test these models
are the latest-type spirals, because they have bulge-less
disks. However, even these objects have rather complex central star
formation histories. About 75\% of late-type disks host distinct
nuclear star clusters \citep{boe02}, most of which have experienced
multiple discrete star formation events in their history
\citep{ros06,wal06}. So far, it is unclear whether the repetitive star 
formation in these nuclear clusters is due to variability of the
gravitational potential (e.g. dissolution/formation of a stellar bar),
the clumpy nature of the molecular gas (e.g. inflow of discrete giant
molecular complexes -- GMCs), or whether nuclear massive star
formation itself is disrupting the gas supply.

The Scd spiral \objectname{IC342} is one of the nearest examples of a
late-type spiral harboring a well-studied nuclear star cluster. The
nuclear cluster has a stellar mass of $\sim 10^7\,\msun$ and its
luminosity is dominated by a recent, 4--30~Myr~old, short-lived star
formation event \citep[after correction for a revised distance of 3.0
Mpc;][]{boe97,boe99,fin07}. Previous studies of the molecular gas in
IC342, with single dish and interferometric millimeter instruments
\citep{tur92,tur93,ish90,dow92,mei01}, lack sufficient spatial
resolution (58--72~pc) to resolve the structure of the GMCs, thus
hampering accurate comparisons with the nuclear star formation. The
spiral geometry of the molecular gas in the central 500pc is due to
the response of the gas to the large-scale stellar bar
\citep{ish90}. A comprehensive study \citep{mei05} of the molecular
chemistry in IC342 has demonstrated that the molecular gas in the
central few hundred pc is subject to a number of different excitation
mechanisms. Recent OVRO observations of the $^{12}$CO(2-1) line
showed evidence for molecular gas being associated with the nuclear
star cluster itself \citep{sch03}.

\section{Observations and absolute astrometry}

The $^{12}$CO(2-1) and HCN(1-0) lines tracing the total and dense
molecular gas content were observed with the IRAM Plateau de Bure
interferometer (PdBI) in 2005 January/February and 2004 January,
respectively. For calibration and mapping, we used the standard IRAM
GILDAS software packages CLIC and MAPPING \citep{gui00}. The
observations resulted in data cubes of $0.67''\times0.51''$ and
$1.64''\times1.29''$ resolution with an rms of 10 and 3.6 mJy/beam for
the CO(2-1) and HCN(1-0) line. Intensity maps were derived using a
3$\sigma$ cut for line emission present over at least two channel of
3.0 (6.0) km/s width. 

For a direct comparison of the (dense) molecular gas to the stellar
light and the ionized gas (as observed with HST), the HST data need to
have an absolute astrometry of better than at least 0.5''. To achieve
this, we first aligned a newly obtained HST Pa$\alpha$ image (PI
Calzetti, PID 11080) to a deep 6cm continuum VLA image (PI Schinnerer)
as both exhibit a very similar geometry at comparable angular
resolution (0.4'' vs. 0.75''). Archival HST F656N and F606W images
(PID 8581, 5446) were then aligned to the Pa$\alpha$ map using a
number of stellar clusters. To obtain the continuum-subtracted
H$\alpha$ map, we subtracted a scaled version of the F606W image from
F656N. No astrometric corrections are required for radio
interferometric data as the observations provide an absolute
astrometric reference frame. This resulted in very well aligned
optical/NIR and radio images, solving the ambiguities that plagued our
previous study which relied on the geometry of the dust lanes
\citep{sch03}.

\section{Gas distribution shaped by star formation}

\subsection{Distribution of the Molecular Gas}

\begin{figure}
\includegraphics[angle=-90.,scale=0.7]{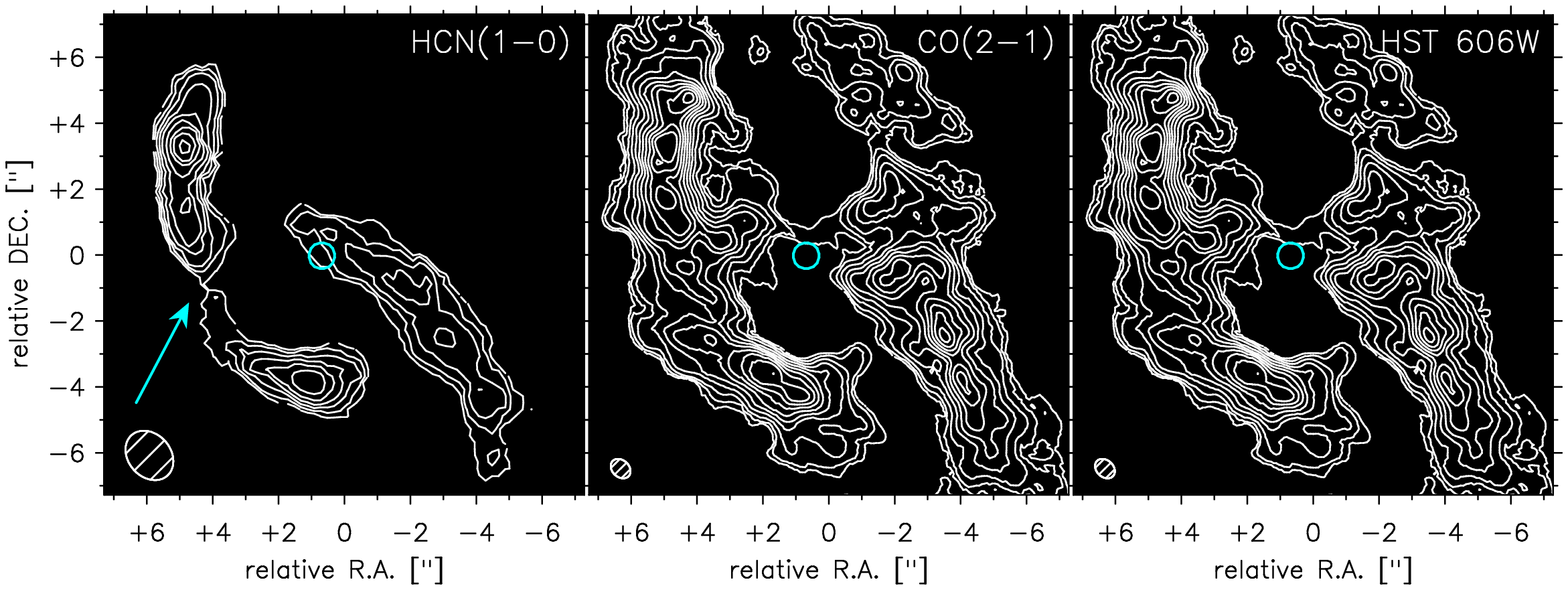}
\caption{Distribution of the cold molecular gas as traced by the
  $^{12}$CO(2-1) line emission ({\it middle}; color and contours
  starting at 0.9 Jy/beam\,km/s in uneven steps) in the central
  225\,pc of the late-type spiral galaxy IC342 at a resolution of
  0.6'' (9.6\,pc). A narrow faint gas bridge connecting the two spiral
  arms can be seen just north of the nuclear cluster (marked by a
  circle in all panels).  The dense molecular gas as traced by the
  HCN(1-0) line emission ({\it left}; color and contours starting at
  0.45 Jy/beam\,km/s in steps of 0.15 Jy/beam\,km/s) shows a very
  similar distribution, although the western spiral arm extends
  further towards the nucleus.  The break (marked by the arrow) about
  4'' east and 1'' south of the center in the eastern spiral arm is
  also more prominent in the HCN map. The molecular gas also coincides
  well with the regions of high extinction in the optical continuum
  image ({\it right}, HST F606W filter in color), while the underlying
  stellar disk is less obscured inside the molecular gas
  'cavity'.\label{fig:sum}}
\end{figure}

\begin{figure}
\epsscale{.80}
\plotone{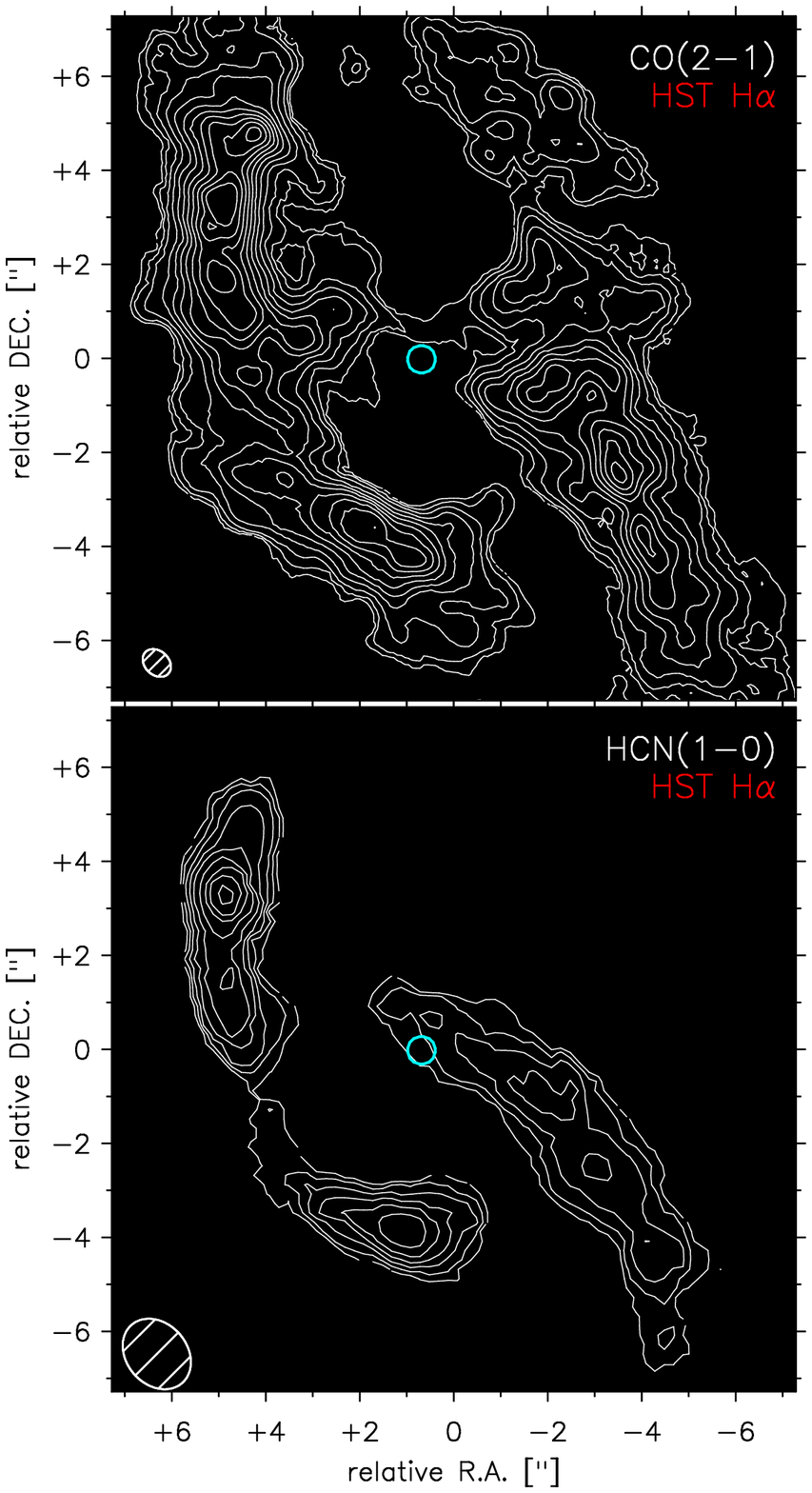}
\caption{ The CO(2-1) ({\it top}) and HCN(1-0) ({\it bottom})
  intensity maps (contours from Fig. \ref{fig:sum}) are overlaid onto
  the continuum-subtracted H$\alpha$ image from HST (color). The
  ionized gas bubble shows a shell-like structure in the Southeast
  that fits well into the 'cavity' of the molecular gas distribution.
  Most of the arcs outlining the bubble appear to have originated from
  the nuclear stellar cluster.
\label{fig:ha}}
\end{figure}

In the new CO(2-1) and HCN(1-0) maps (Fig. \ref{fig:sum}), the Eastern
spiral arm exhibits a prominent break at $3^{h}46^{m}49.0^{s}$ and
$+68^{o}5'45.5''$ (J2000), about 4'' east and 1'' South of the galaxy
center. At this break the Eastern spiral arm stops continuing inward
and stays at a roughly constant distance from the nucleus whereas the
Western spiral arm reaches close (within 1'') to the nucleus itself.
The nuclear CO(2-1) gas clump identified earlier \citep{sch03} is now
resolved into a faint, narrow gas lane running East-West just North of
the nucleus. In addition, the Southern portion of the Eastern spiral
arm exhibits a sharp edge on the side facing the nucleus. The geometry
of the HCN emission closely resembles that of the CO(2-1), except for
the Western spiral arm which extends further towards the nucleus in
HCN while the emission North of it is too diffuse to be detected in
HCN. The central 'cavity' in the molecular gas distribution is
displaced from the nucleus to the Southeast and offers a relatively
unobscured view onto the stellar disk, as demonstrated by the
comparison to the optical continuum (Fig. \ref{fig:sum}): the stellar
disk is substantially fainter wherever molecular gas is present. Both
CO(2-1) and HCN emission trace very well the prominent dust lanes in
the central 250pc. This dramatic change in extinction causes an
elongation of the isophotes even at NIR wavelengths, which might be
misinterpreted as a small-scale bar. A comparison of the CO(2-1) and
H$\alpha$ distribution (Fig. \ref{fig:ha}) shows that the ionized gas
fills the central 'cavity' of the molecular gas. In particular,
the H$\alpha$ shells (or arcs) delineating the Southeast edge of the
cavity suggest an ionized outflow that coincides with the inner edge
of the Southeastern CO(2-1) spiral arm. Due to IC342's almost face-on
orientation, the gas motions are difficult to interpret, as we start
to observe a blend of several smaller molecular clouds plus streaming
motions due to the stellar bar and a mix of the diffuse and dense gas
component in the CO(2-1) emission.

\subsection{Impact of Nuclear Star Formation on a Bar-Induced Gas Flow}

While the overall molecular gas distribution in the center of
IC342 is governed by the gravitational influence of a large-scale
stellar bar \citep{ish90} \citep[also as evidenced by the presence of
shocks seen in CH$_{3}$OH and SiO,][]{mei05,use06}, the asymmetries in
the molecular gas properties within the inner 50 pc can be explained
by the impact of nuclear star formation onto this gas flow. The
comparison of the molecular gas distribution to the H$\alpha$ image
suggests that the expanding ionized gas is 'running' into the
molecular gas, particularly in the Southeastern portion of the spiral
arm (hereafter: segment). As we will show in \S \ref{sec:feed},
the displacement of this arm segment is in agreement with mechanical
energy input into the total and dense molecular gas, leading to it
being compressed and pushed out, or at least 'halted' by the pressure
of the ionized gas.

\cite{mei05} have shown that this part of the spiral (GMC--A) is the
one most excited by UV photons and more diffuse than the rest of the
dense GMCs. Radiative excitation by the nucleus can be excluded as it
is not an AGN or a UV source \citep{boe99}. The discontinuity in the
Eastern spiral arm as well as the different chemical properties of
this arm segment argue against a simple offset of the gas spiral from
the dynamical center. Taken together, this suggests an impact from
stellar winds and supernova explosions on the morphology of
circumnuclear molecular gas as well as altering its chemical
composition.

Some evidence for a corresponding discontinuity along the Northern
portion of the western arm (thin dotted line in Fig. \ref{fig:sketch})
is observed in CO(2--1). However the weak H$\alpha$ emission, the lack
of an altered chemistry and the HCN morphology makes the case for
direct impact of stellar winds and SNe less clear here. The exact
reasons for the asymmetry manifested by the outflow from the nuclear
starburst (as seen in the H$\alpha$ line) are unclear, but a such
behavior is not unusual for large-scale outflows from central star
formation \citep[e.g. M82;][]{ohy02}.  Potentially the Northern
portion of the Western spiral arm might lack these signs because it is
composed of more dispersed molecular material. In addition, the higher
extinction on the near-side (Western side) of the galaxy can cause
some differences as well.

Although the dynamical timescale in the nuclear region is very small
($\sim$ 10\,Myr), the spiral pattern itself is rotating slower. As
nuclear spirals can easily be excited by large-scale bars
\citep[e.g.,][]{mac04,eng00}, we can use the pattern speed
of the stellar bar of $\Omega_{bar}$ = 0.4 km/s/'' \citep{cro00}. At a
radius of 3.3'' (50pc) this corresponds to a velocity of 8\,km/s, thus
one revolution of the spiral pattern itself takes about 40\,Myr. For
a 10 Myr old starburst the spiral has moved by a quarter. This is
consistent with the arm segment which covers about a quarter of a circle
and also the age for the most recent star formation event in the
nuclear cluster.

To summarize, the ionizing outflow produced by stellar winds and
supernovae from the recent nuclear star forming event appears to have
changed the location of the Southern part of the Eastern spiral arm,
while the Western spiral arm appears less effected. If the gas
flow along the western arm is less disturbed by the nuclear bubble,
this may explain why the youngest star formation episode is also
asymmetrically stronger towards the western arm
\citep[e.g.][]{tsa06}.

\section{Evidence for Feedback}\label{sec:feed}

We now investigate whether the timescale and required energy are
indeed compatible with the scenario outlined above. The dislocation is
either due to an actual movement of individual molecular clouds against
the gravitational potential or the ionized gas is acting as a barrier
altering the orbits of the molecular gas. In both scenarios, some kind
of force is required to counterbalance the gravitational potential,
thus the following estimates should be valid in either case.

The deprojected distance from the nuclear cluster to the
H$\alpha$ rim is about 3.3" or 50pc. A shell with a typical expansion
velocity of 20-40 km/s \citep{mar98} can travel this distance in about
(2.4 - 1.2) Myr, which is less than the age of the last nuclear star
formation episode (4-30 Myr). To first order, it is therefore possible
that the H$\alpha$ shells were produced in this event. Assuming that
the Southeastern arm segment has been `pushed' to a larger radius, we
can derive a rough estimate for the required cumulative energy using
the mass of the molecular gas and reasonable assumptions about the
gravitational potential.

We assume a virial gas mass of $1.6\times10^6\,\msun$ for the
Southeastern segment \citep{mei01}. To estimate the
kinetic and potential energy required, we use the rotation curve of
\cite{tur92} after correcting for the new distance to estimate
velocities of v(r$_2$=50pc)$\approx$25 km/s and
v(r$_1$=22pc)$\approx$12 km/s. A total energy of $\rm
E_{tot}=E_{kin}+E_{pot}=3\cdot
E_{kin}=3\cdot\,m_{GMC-A}\,[v(r_2)^2\,-\,v(r_1)^2]/2\cong
2.2\times10^{52} erg$ is required to move the molecular gas mass from a
radius of 18pc (the location of the unaffected Western spiral arm) to
50pc, the current distance of the Southeastern segment
\citep[inclination effects are neglected given the low inclination of
$\sim31^o$,][]{cro00}.

A cluster with a stellar mass of $1\times10^7\,\msun$ releases a
mechanical luminosity of $\rm 3\times10^{41}\,erg\,s^{-1}$ at an age
of 4-30 Myr \citep[assuming an instantaneous burst with solar
metallicity; see STARBURST99 models,][]{lei99}. Since the
flux-calibrated continuum-subtracted HST Pa$\alpha$ image suffers less
dust extinction than the H$\alpha$ image (Fig. \ref{fig:ha}), we use
the former for a consistency check. The Pa$\alpha$ line flux of the
ionized gas bubble ($\rm \sim 7\times10^{-13}\,erg\,s^{-1}cm^{-2}$)
can be converted into a H$\alpha$ luminosity of $\rm
5.4\times10^{39}erg\,s^{-1}$ using a H$\alpha$/Pa$\alpha$ ratio of
7.82 appropriate for a metal-rich region (without extinction
correction). Assuming that about 3\% of the mechanical energy is
converted into H$\alpha$ luminosity
\citep{bin85} and a filling factor of about 0.5 for the bubble (from
the flux present inside the bubble), this translates into a mechanical
luminosity of about $\rm 3.6\times10^{41}erg\,s^{-1}$. This is
consistent with the above value derived from the cluster mass and age.

Thus over the course of the derived travel time of 2 Myrs a total
mechanical energy of $\rm 2\times10^{55}erg$ will be produced.
However, this mechanical energy is distributed over the entire sphere
and subject to radiative losses. Assuming a filling factor of only 3\%
for GMC-A and radiative losses of 90\%, plus taking into account that
likely only a fraction of 20\% of the nuclear stellar mass was
produced in the last event, an energy of $\rm 1.2\times10^{52}erg$ is
expected to be available for mechanical interaction. Even this rather
conservative estimate for the available mechanical energy is
sufficient to explain the gas morphology. We conclude that, based on
time and energetic arguments, mechanical feedback from nuclear star
formation is a plausible explanation for the observed molecular gas
distribution.

\section{Self-Regulated Fueling}

\begin{figure}
\epsscale{0.6}
\plotone{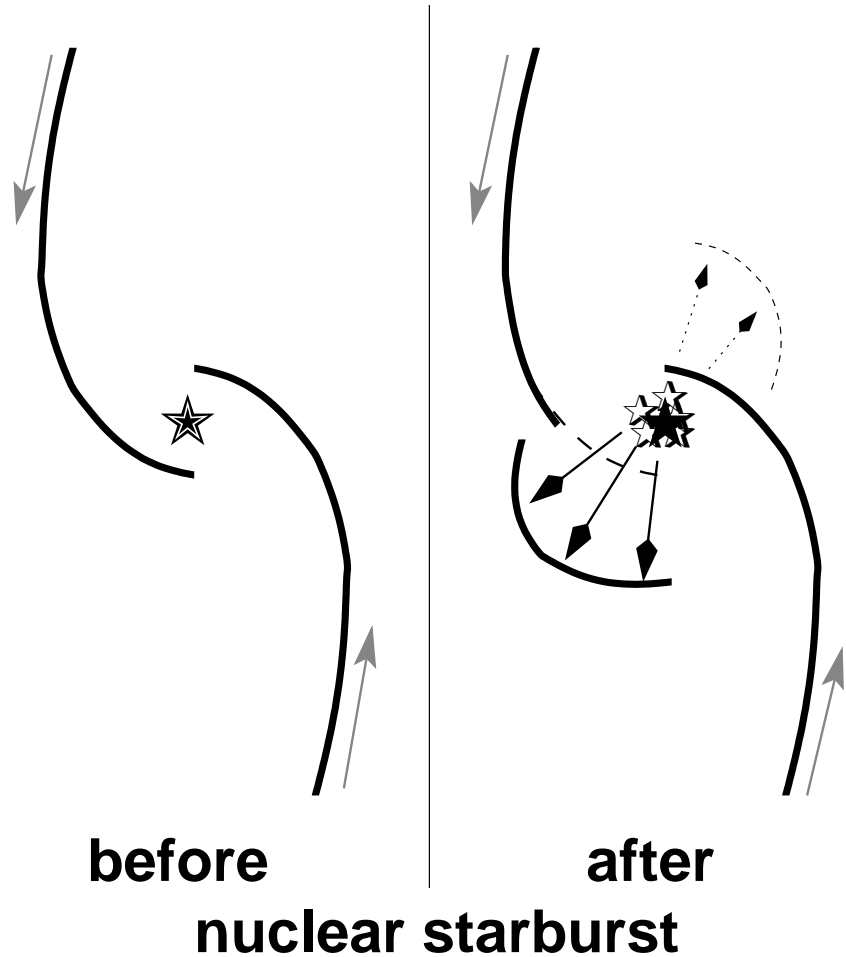}
\caption{
Schematic figure outlining the proposed scenario for the
self-regulation of the fueling of the nuclear stellar cluster in
IC342. {\it Left:} In the absence of massive star formation the
molecular gas moves towards the nucleus (star) along the gas lanes
(black arcs) due to the gravitational influence of a large-scale
stellar bar. {\it Right:} The mechanical energy (indicated by black
arrows) released by stellar winds and supernovae of the recent massive
star formation in the nucleus is sufficient to alter the flow of the
molecular gas (from the dashed to the solid position), so that
molecular gas from the Eastern spiral arm can no longer reach close to
the nucleus, thus significantly reducing the fuel available for future
star formation events. The effect on the Western arm is less dramatic
(thin dotted line and arrows). After the stellar winds and SN
explosions have ceased, it is expected that the force of the
gravitational bar potential takes over again and the gas flow will be
similar to the situation shown in the left panel.
\label{fig:sketch}}
\end{figure}

The following picture for the gas flow in IC342 emerges (Fig.
\ref{fig:sketch}): The large-scale stellar bar moves gas toward the
center via two gas spiral arms. The stellar winds and supernova shocks
released by the most recent nuclear star formation event have
significantly altered the path of the molecular gas towards the center
on the last few 10pc, thus currently preventing efficient fueling of
the nucleus. This is the first time strong evidence has been found
for mechanical feedback of nuclear star formation onto the (disk) gas
{\it flow} in an extragalactic nucleus. As the large-scale stellar
bar continues to move gas towards the center, it is likely that the
gas resumes its old path once the mechanical energy has been
dispersed.  Presumably, the nucleus will then again collect molecular
gas until the next star formation episode. This feedback between
nuclear activity and fueling efficiency appears to self-regulate the
rate of nuclear star formation and offers a natural explanation for
the repetitive star formation found in nuclear star clusters
\citep{ros06,wal06}.

If correct, this scenario implies that models for the gas fueling
mechanism (over the innermost 1-100\,pc) cannot rely on the shape of
the gravitational potential alone but need to take the effect of
mechanical energy released by nuclear activity onto the gas flow into
account. The impact of this feedback process will be strongly
variable in time and will critically depend on the amount of energy
released as well as its geometry. Although mechanical feedback has
long been suspected to be important for galactic nuclei, so far no
observations have existed to directly support this. In a broader
sense, these results also have implications for the evolution of any
central massive object (be it a stellar cluster or a black hole) that
requires occasional fueling. Nuclear activity appears to have the
potential to significantly reduce or even (temporarily) shut off
fueling, thus self-regulating the growth of the central mass.

\acknowledgments

We thank J. Knapen and K. Wada for their constructive criticism. ES
would like to thank the IRAM staff for their help with the data
reduction.

{\it Facilities:} \facility{IRAM PdBI}, \facility{HST (WFPC2,NICMOS)}.

\clearpage

\end{document}